\documentclass[prb,preprintnumbers,superscriptaddress,amsmath,amssymb,floatfix,aps,twocolumn]{revtex4}
\usepackage{graphicx}

\usepackage{lipsum} 
\usepackage{comment}
\usepackage{lineno}

\usepackage{xcolor}

\newcommand{\bluemark}[1] {\color{black}#1\color{black}\normalsize}

\usepackage{ulem}
\newcommand{\ket}[1]{\left\lvert #1 \right\rangle}

\usepackage{afterpage}

\usepackage[version=4]{mhchem}

\usepackage{hyperref}
\usepackage{amsmath}
\usepackage{amsbsy}
\usepackage{float}
\usepackage[utf8]{inputenc} 

\begin{document}

\title{Dual Topology as a Fingerprint of Relativistic Altermagnetism in AgF$_2$ Monolayer}

\date{\today} 

\author{J. W. Gonz\'alez}
\email{jhon.gonzalez@uantof.cl}
\affiliation{Departamento de Física, Universidad de Antofagasta, Av. Angamos 601, Casilla 170, Antofagasta, Chile}

\author{R. A. Gallardo}
\affiliation{Departamento de F\'{i}sica, Universidad 
T\'{e}cnica Federico Santa Mar\'{i}a, Casilla Postal 
110V, Valpara\'{i}so, Chile.}

\author{N. Vidal-Silva}
\affiliation{Departamento de Ciencias F\'{i}sicas, Universidad de La Frontera, Casilla 54-D, Temuco, Chile}

\author{A. M. León}
\email{andrea.leon@uchile.cl}
\affiliation{Departamento de Física, Facultad de Ciencias, Universidad de Chile, Casilla 653, Santiago, Chile.}

\begin{abstract}
Altermagnets have emerged as a fertile ground for quantum phenomena, but topological phases unifying different quasiparticles remain largely unexplored. Here, we demonstrate that monolayer AgF$_2$ hosts a dual topological state, driven by a single ferroelastic distortion. This polar transition breaks inversion symmetry and unleashes relativistic spin-orbit effects, simultaneously imparting non-trivial topology to electrons and magnons. The result is valence bands with opposite Chern numbers, $C^E=\pm3$, and a magnon spectrum with a full topological gap and chiral bands, $C^M=\pm1$. This work realizes topological altermagnonics in a tangible material platform, with a clear experimental fingerprint in the transverse thermal Hall effect. The coexistence of fermionic and bosonic topology in AgF$_2$ opens new directions for designing intrinsically hybrid quantum matter.
\end{abstract}

\maketitle

\section{Introduction}
Topological phases of matter, first identified in electronic systems, have recently been extended to bosonic quasiparticles such as magnons, the quantized spin waves in magnetic insulators~\cite{zhuo2025topological,mcclarty2022topological, aguilera2020topological, vidal2022time}. While their study has been fruitful in ferromagnets, realizing such features in conventional collinear antiferromagnets (AFMs) is challenging, as their magnon branches are typically degenerate due to high crystal symmetry~\cite{Khatua2025}. Altermagnetism (AM), a newly recognized class of collinearly compensated magnets, circumvents this challenge via a unique rotational symmetry that intrinsically lifts the magnon degeneracy, providing an ideal platform for realizing topological phases~\cite{Smejkal2022_Review, hoyer2025altermagnetic, jungwirth2025altermagnetism}.
\bluemark{Within the recently established parity framework of unconventional magnets~\cite{fukaya2025superconducting}, altermagnets belong to the even-parity class, where spin splitting is symmetric under momentum inversion, $\Delta E(\mathbf{k})=\Delta E(-\mathbf{k})$~\cite{Smejkal2022_Review}. This contrasts with odd-parity $p$-wave magnets, which arise from non-collinear spin textures and feature $\Delta E(\mathbf{k})=-\Delta E(-\mathbf{k})$~\cite{hellenes2023p, song2025electrical}. This classification is essential for contextualizing how relativistic interactions can hybridize odd- and even-parity characteristics, a mechanism we uncover in monolayer AgF$_2$~\cite{sukhachov2025coexistence}.}

\begin{figure*}[!]
\centering
\includegraphics[clip,width=0.7\textwidth,angle=0]{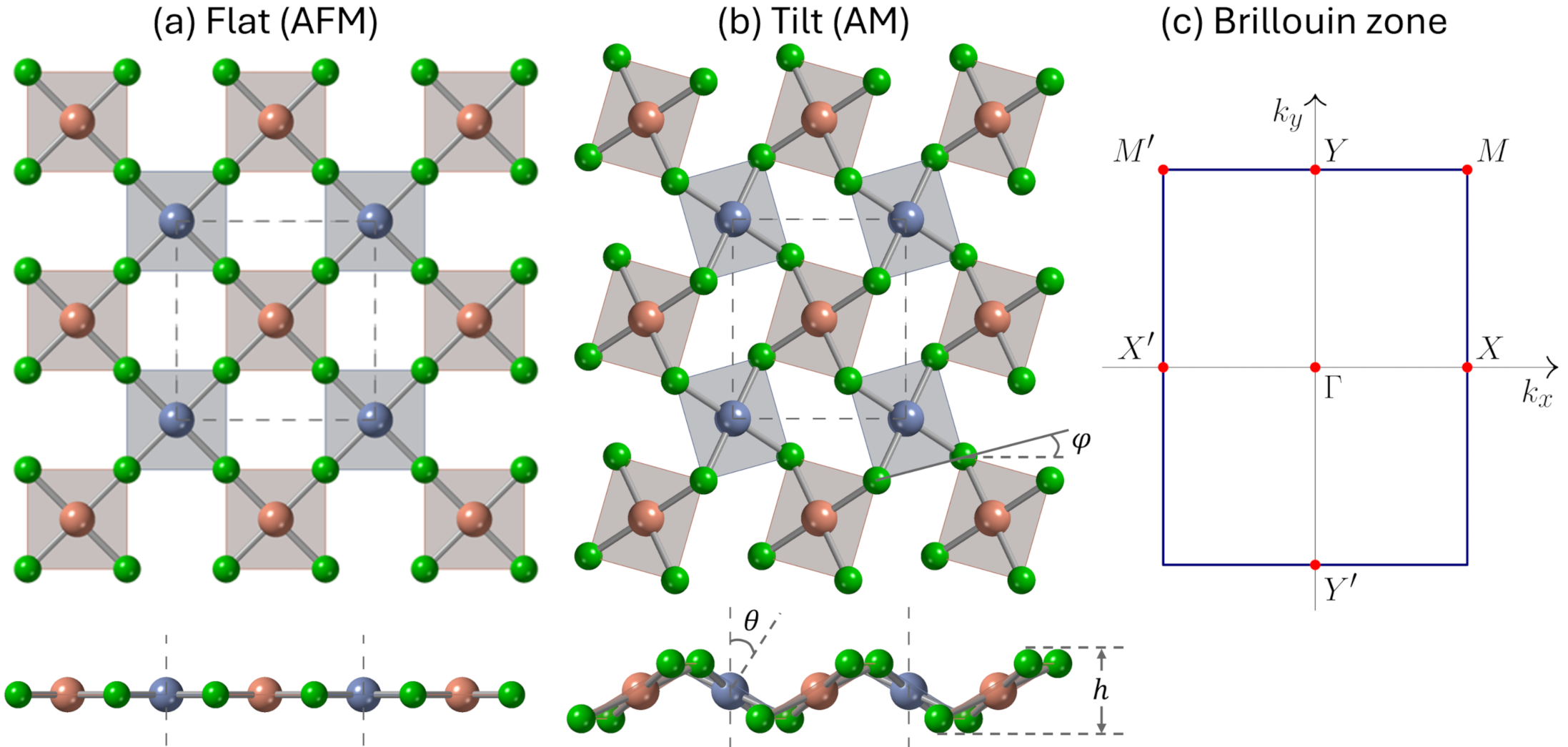} 
\caption{%
Structural representations of monolayer AgF$_2$. Green spheres represent fluorine (F) atoms. 
Blue and red spheres represent silver (Ag) atoms with opposite magnetic moment projections (positive for blue and negative for red).  
\textbf{(a)}~Flat, high-symmetry reference structure with $a = b$, corresponding to the tetragonal space group $P4/mmm$ (No.\,123), used as a non-polar reference phase. 
\textbf{(b)}~Tilted ground-state geometry with characteristic octahedral tilting and polar distortions, belonging to the monoclinic space group $P2_1$ (No.\,4). Tilting angles $\theta$, $\varphi$, and the vertical displacement $h$ illustrate the broken symmetries. 
\textbf{(c)}~First Brillouin zone of the distorted geometry, with high-symmetry $k$-points labeled.}
\label{fig:scheme}
\end{figure*}

Silver fluorides, particularly AgF$_2$, represent a compelling frontier in correlated electron systems because they are the closest known analogs of the cuprates~\cite{gawraczynski2019silver,bandaru2021fate}. This has fueled extensive theoretical predictions of unconventional $d$-wave superconductivity upon doping~\cite{liu2022silver}, akin to their cuprate counterparts. Yet this scenario faces severe experimental hurdles: superconductivity in AgF$_2$ remains elusive, as carrier doping is strongly frustrated by phase separation and polaronic self-trapping~\cite{grzelak2021defect, bandaru2021fate}. More recently, the discovery of altermagnetism has opened an entirely different pathway. Theory now predicts that altermagnets can host unconventional spin-triplet ($p$-wave or $f$-wave) and even topological superconductivity~\cite{hong2025unconventional}, a mechanism distinct from the cuprate paradigm. Our work lies at the intersection of these two routes. We demonstrate that monolayer AgF$_2$ stabilizes a polar altermagnetic ground state, a phase where the key ingredients for both scenarios coexist. Crucially, we show that this state is not a passive starting point but a distinct form of emergent quantum matter: a dual topological phase with intrinsically coupled electronic and magnonic orders.

In this work, we identify AgF$_2$ as a compelling candidate in which a structural distortion drives the system from a centrosymmetric, topologically trivial AFM to a polar, ferroelastic altermagnetic ground state~\cite{Peng2025_Ferroelastic}. This transition not only activates a non-relativistic altermagnetic spin splitting but, upon inclusion of spin-orbit coupling (SOC), also stabilizes a striking electronic topological phase where the two highest valence bands carry opposite Chern numbers, $C^E=\pm3$. In parallel, the same distortion enhances magnetic interactions, generating a robust Dzyaloshinskii-Moriya term (DMI) that opens a full topological gap in the magnon spectrum and yields non-degenerate bands with $C^M=\pm1$ \cite{vidal2022time}. These topological magnons constitute a direct realization of altermagnonics, giving rise to a finite transverse thermal Hall conductivity that provides a clear experimental hallmark. Our results, therefore, unveil an unprecedented platform where a single structural switch entangles electronic and magnonic topology, charting a route toward multifunctional quantum materials that intrinsically couple fermionic and bosonic quasiparticles.

\bluemark{Our findings establish monolayer AgF$_2$ as a unique platform where a single, symmetry-lowering mechanism imparts non-trivial topology to both electronic and magnonic excitations. This discovery introduces a symmetry-based route to engineer coupled fermionic and bosonic topological transport in collinear magnets. More broadly, our results suggest that planar two-dimensional transition-metal halides~\cite{botana2019electronic,fukuda2021structure}, MX$_2$ (M = transition metal; X = halide), constitute a versatile class of quantum materials where structural distortions and relativistic interactions conspire to generate topological phases across multiple quasiparticle sectors. Taken together, these insights establish 2D halide magnets as a unifying platform where intertwined topological fermions and bosons may emerge as a generic feature, opening avenues toward device concepts that harness electron-magnon interconversion and hybrid quantum transport.}

\section{Methodology}
\bluemark{
We perform first-principles calculations with the Vienna Ab-initio Simulation Package (VASP)~\cite{VASP} using the projector augmented-wave  method~\cite{kresse1999ultrasoft}. Exchange-correlation effects are described within the generalized gradient approximation in the Perdew-Burke-Ernzerhof  parametrization~\cite{PBE}. We use a kinetic-energy cutoff of 450~eV for the plane-wave basis set, above the recommended values of the employed PAW pseudopotentials.  
To extract the magnetic exchange coupling constants, additional calculations are performed within a localized basis-set DFT framework using the OpenMX package~\cite{ozaki2004numerical,ozaki2005efficient} within equivalent parameters.}  

\bluemark{
We model the AgF$_2$ monolayer in a slab geometry with a vacuum region of at least 15~\AA\ along the out-of-plane direction to prevent spurious interlayer interactions. The Brillouin zone is sampled with Monkhorst-Pack meshes of $10\times9\times1$ for structural relaxations and $20\times18\times1$ for static self-consistent calculations, corresponding to $k$-point spacings of approximately $0.02$ and $0.01$ $2\pi/$\AA{}, respectively~\cite{Monkorst}. We use Methfessel-Paxton smearing with a width of 0.05~eV during ionic relaxations. Atomic positions are relaxed until the residual forces on each atom are smaller than $10^{-3}$~eV/\AA, with electronic self-consistency reached within $10^{-6}$~eV.  
To account for strong on-site Coulomb interactions of Ag $4d$ states, we apply the rotationally invariant DFT+$U$ approach in the Dudarev formalism~\cite{dudarev1998electron}, with an effective parameter $U_{\mathrm{eff}} = 4$~eV. This choice, consistent with earlier studies of Ag(II) fluorides~\cite{domanski2021theoretical,gawraczynski2019silver}, reproduces key experimental observables such as the insulating band gap and the magnitude of local magnetic moments in AgF$_2$.  
}


\section{Electronic properties}

The physical properties of monolayer AgF$_2$ are intrinsically linked to its crystal structure. This work studies two key structural polymorphs, depicted in Figure~\ref{fig:scheme}: (a) a high-symmetry, flat geometry and (b) a distorted, tilted ground-state geometry. These structures differ fundamentally in their point group symmetry, fluorine atom arrangement, and, most critically, their inversion properties. The subsequent sections demonstrate how these crystallographic distinctions give rise to profoundly different electronic and magnetic behaviors, establishing a direct connection between the crystal geometry and the emergent topological phenomena.

\begin{figure*}[!]
\centering
\includegraphics[clip,width=0.7\textwidth,angle=0]{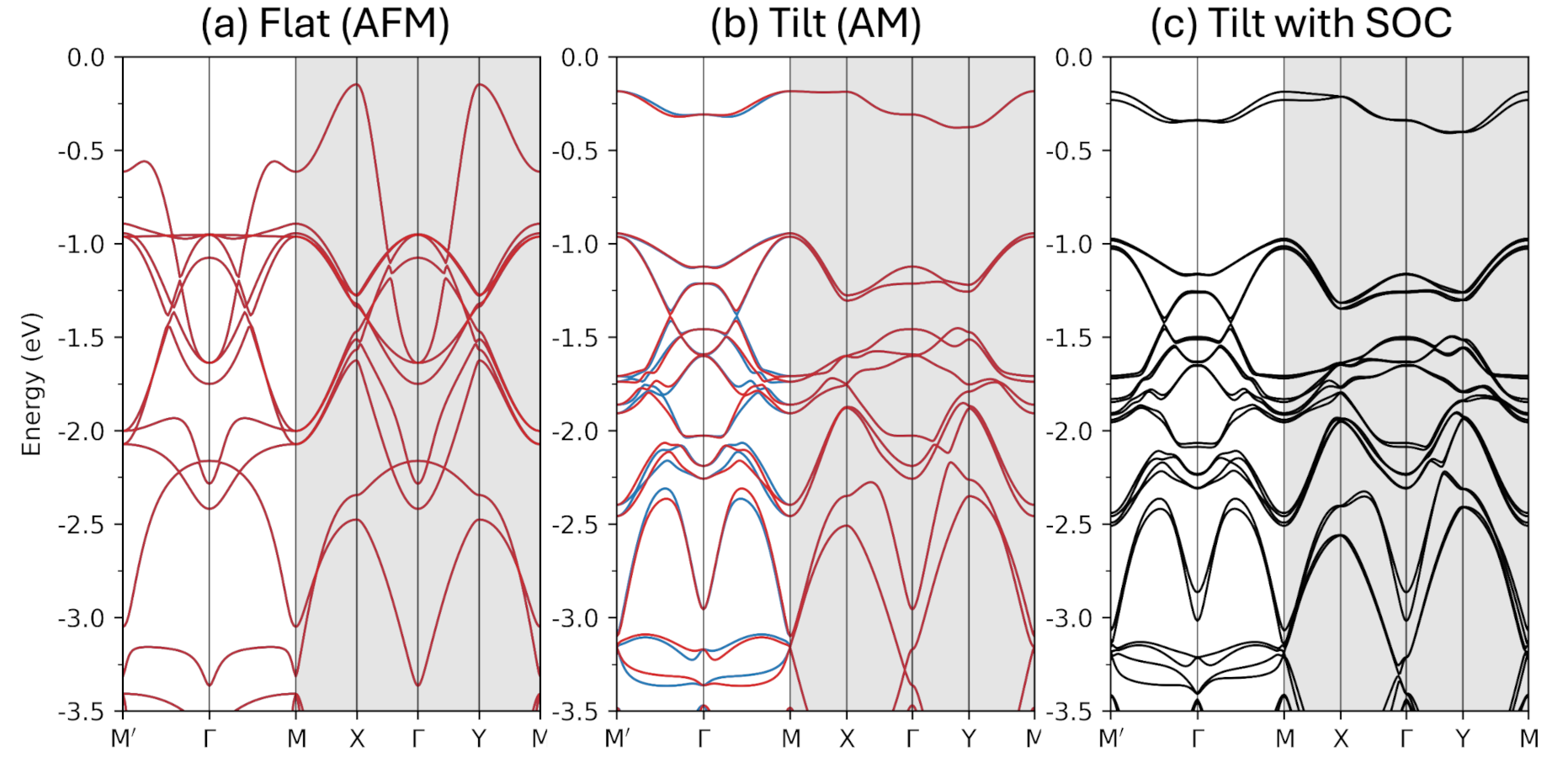} 
\caption{%
Electronic band structures of monolayer AgF$_2$ along the $M'$-$\Gamma$-$M$-$X$-$\Gamma$-$Y$-$M$ high-symmetry path.
\textbf{(a)}~Flat antiferromagnetic (AFM) phase, corresponding to the tetragonal space group $P4/mmm$ (No.\,123), displays fully spin-degenerate bands.
\textbf{(b)}~Altermagnetic (AM) ground state with octahedral tilting, belonging to the monoclinic space group $P2_1$ (No.\,4), exhibits momentum-dependent spin splitting.
Red and blue curves indicate spin-resolved bands within the AM window.
\textbf{(c)}~AM phase with spin-orbit coupling included, lifting the remaining degeneracies and inducing pronounced splittings at high-symmetry points, particularly near $M$ and $M'$.}
\label{fig:bands-flat-tilt}
\end{figure*}

\subsection{Geometry: From centrosymmetric\,AFM to polar\,AM \label{sec:flat-vs-tilt}}

The flat polymorph, shown in Fig.~\ref{fig:scheme}(a), provides a high-symmetry structural reference. It possesses a tetragonal lattice described by the centrosymmetric, non-polar space group $P4/mmm$ (No.~123), with all atoms confined to a single plane. The AgF$_4$ square motifs are perfectly flat and undistorted. Consequently, the two Ag sublattices are crystallographically equivalent, related by an inversion center. Our DFT calculations confirm this phase is metastable, with a magnetic moment of 0.453 $\mu_B$ lies approximately 0.2~eV per formula unit above the ground state. In this configuration, the inversion symmetry connects the two Ag sites. 
In contrast, the actual structural ground state, depicted in Fig.~\ref{fig:scheme}(b), adopts the polar monoclinic space group $P2_1$ (No.~4). Here, the lattice undergoes a cooperative distortion involving the tilting and buckling of the AgF$_4$ units, which breaks both the fourfold rotational and, most critically, the inversion symmetries. The resulting relaxed unit cell is anisotropic, with a calculated in-plane aspect ratio of $a/b \approx 0.87$ and a magnetic moment of 0.528 $\mu_B$. This distortion is quantified by three primary parameters: an out-of-plane tilt $|\theta|= 24.1^\circ$, an in-plane rotation $|\varphi_{21}| = 14.4^\circ$, and a vertical buckling height $h = 1.49~\text{\AA}$. The deformation also manifests as a slight asymmetry in the F-F distances within the AgF$_4$ units (2.947~\text{\AA} and 2.924~\text{\AA}). The critical consequence of this complex deformation is that it renders the two Ag sublattices crystallographically inequivalent by removing the inversion center that previously related them. 

\begin{figure*}[!]
\centering
\includegraphics[clip,width=0.7\textwidth,angle=0]{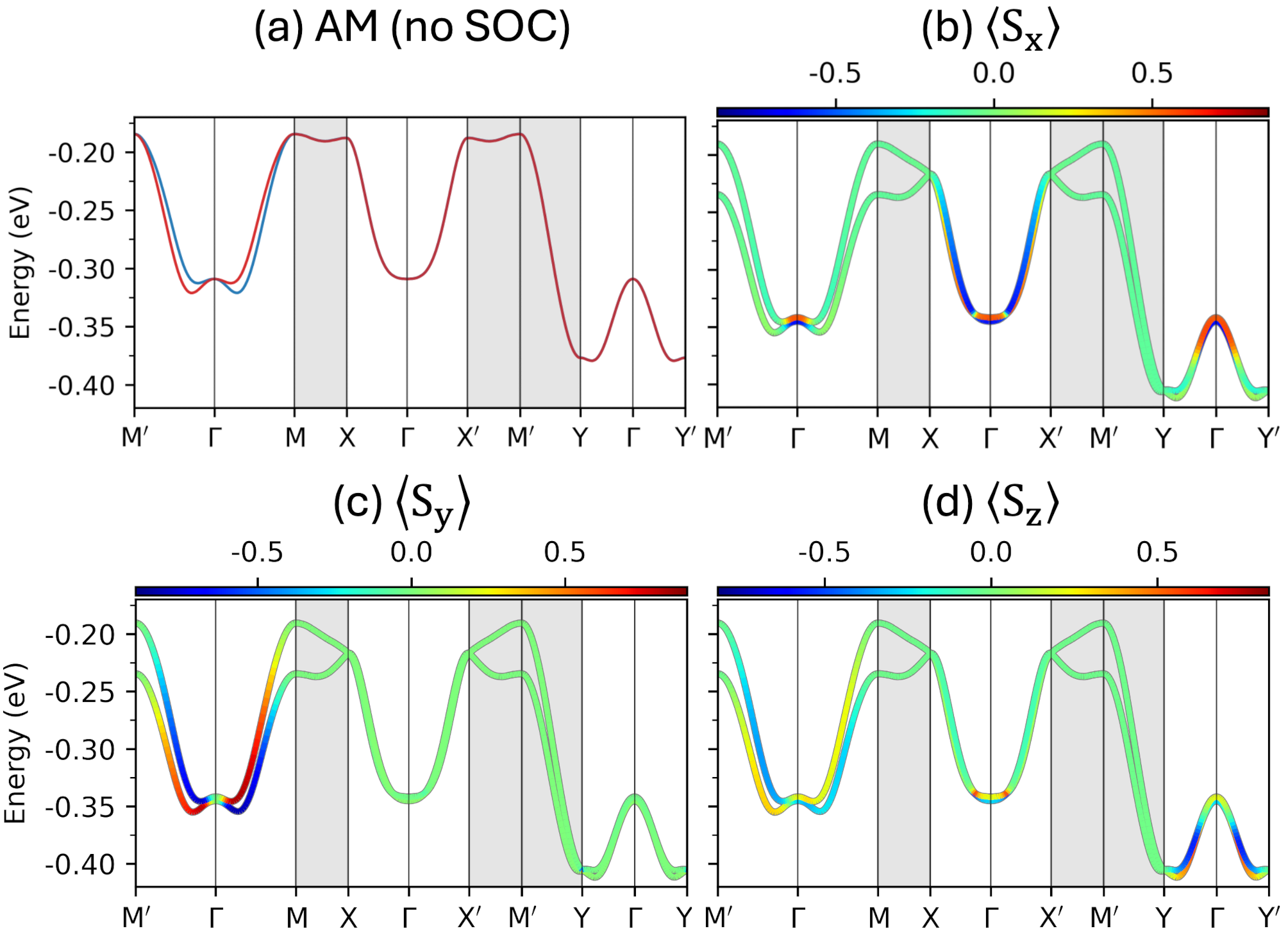} 
\caption{%
Spin-resolved electronic band structures of monolayer AgF$_2$, focused on the top valence bands V1 and V2.
The $k$-path is extended to include the $\Gamma$-$\mathrm{X}'$ and $\Gamma$-$\mathrm{Y}'$ directions to fully capture the symmetry of the Brillouin zone.
\textbf{(a)}~Altermagnetic (AM) ground state without spin-orbit coupling (SOC), exhibiting non-relativistic spin splitting.
\textbf{(b-d)}~Band structures with SOC included considering magnetic moments aligned along $y$-axis, colored by the spin expectation value along each Cartesian axis:
\textbf{(b)}~$\langle S_x \rangle$,
\textbf{(c)}~$\langle S_y \rangle$, and
\textbf{(d)}~$\langle S_z \rangle$.}
\label{fig:bands-soc}
\end{figure*}

\bluemark{
To establish the origin of altermagnetism in AgF$_2$, we performed a symmetry analysis using \textsc{FINDSYM}~\cite{stokes2005findsym,FINDSYM} with strict tolerances. The distorted phase is identified as $P2_1$ (No.~4), a non-centrosymmetric polar space group arising from octahedral tilting and buckling of the tetragonal $P4/mmm$ parent. This distortion breaks inversion and fourfold rotational symmetry, leaving a nonsymmorphic screw axis ($2_1$) as the key operation. The $2_1$ symmetry maps the two Ag sublattices onto each other without restoring inversion, thereby fulfilling the defining condition for altermagnetism~\cite{Smejkal2022_Review}. Crucially, this symmetry reduction enables spin-orbit coupling to generate DMI interactions, which are strictly forbidden in centrosymmetric $P4/mmm$, but are essential for stabilizing the topological magnon bands of the altermagnetic state.
}

The relationship between crystal symmetry and electronic structure is directly reflected in the band dispersions of the two polymorphs [Fig.~\ref{fig:scheme}(c)]. In the high-symmetry flat phase, the bands remain fully spin-degenerate throughout the Brillouin zone [Fig.~\ref{fig:bands-flat-tilt}(a)], consistent with centrosymmetry and time-reversal symmetry in a conventional, trivial antiferromagnet. The system exhibits a direct gap of $\sim$1.8~eV, with the valence-band maximum at $M$ and the conduction-band minimum at $\Gamma$.  
By contrast, the reduced symmetry of the tilted ground state permits a non-relativistic, momentum-dependent spin splitting, a defining feature of altermagnetism. As shown in Fig.~\ref{fig:bands-flat-tilt}(b), this splitting is highly anisotropic: it reaches $\sim$100~meV along $M'$--$\Gamma$--$M$, vanishes along $\Gamma$--$Y$, and is strongly suppressed along $\Gamma$--$X$. This pattern reflects the screw symmetry of the $P2_1$ space group: along $\Gamma$--$Y$, the operation $\{C_{2b} \mid \tfrac{1}{2} \mathbf{b}\}$ preserves local degeneracy, while along generic directions such as $M'$--$\Gamma$--$M$ no such protection exists\cite{Qayyum2024,Sheoran2024}. The resulting exchange-driven, $d$-wave–like anisotropic spin texture is a hallmark of altermagnetic phases, and originates directly from the symmetry-lowering structural distortion.

\subsection{AM+SOC: Electronic spin-orbit effects \label{soc_elect}}

The inclusion of relativistic spin–orbit coupling (SOC) qualitatively enriches the tilted altermagnetic (AM) phase of AgF$_2$. SOC lifts residual degeneracies at symmetry-protected crossings, most notably near the $M$ and $M'$ points, where it opens gaps of up to 20~meV [Fig.~\ref{fig:bands-flat-tilt}(c)]. This behavior reflects the absence of inversion and screw symmetries in the $P2_1'$ magnetic space group, consistent with the group-theoretical framework of altermagnetism~\cite{Smejkal2022_Review,Sheoran2024}, while a detailed symmetry analysis is presented in the Supplemental Material~\cite{MAXMAGN,aroyo2006bilbao,bilbao}.  

Beyond its role in band degeneracies, SOC also determines the magnetic ground state. Our calculations yield a magnetic anisotropy energy (MAE) of $\sim$0.2~meV/f.u., favoring in-plane spins, and a weak ferromagnetic canting that produces a small net moment of $\sim$0.1~$\mu_B$. This canting, symmetry-allowed in the polar $P2_1$ phase and driven by Dzyaloshinskii–Moriya interactions~\cite{moriya1960anisotropic}, has direct experimental implications: it couples AM order to external magnetic fields, induces nonreciprocal magnon dispersion relevant for magnonic circuits~\cite{jungwirth2025altermagnetism}, and provides accessible probes through spin-torque or magneto-optical measurements.  
Finally, SOC seeds an additional odd-parity $p$-wave response on top of the dominant $d$-wave altermagnetic splitting. Such hybridization, absent in the non-relativistic limit, broadens the symmetry-breaking landscape of relativistic altermagnets and enables anisotropic transport phenomena including nonreciprocal conductivity, directional dichroism, and spin-galvanic couplings~\cite{fukaya2025superconducting,wang2025two,leon2025strain}.

 \begin{figure*}[!]
\centering
\includegraphics[clip,width=0.66\textwidth,angle=0]{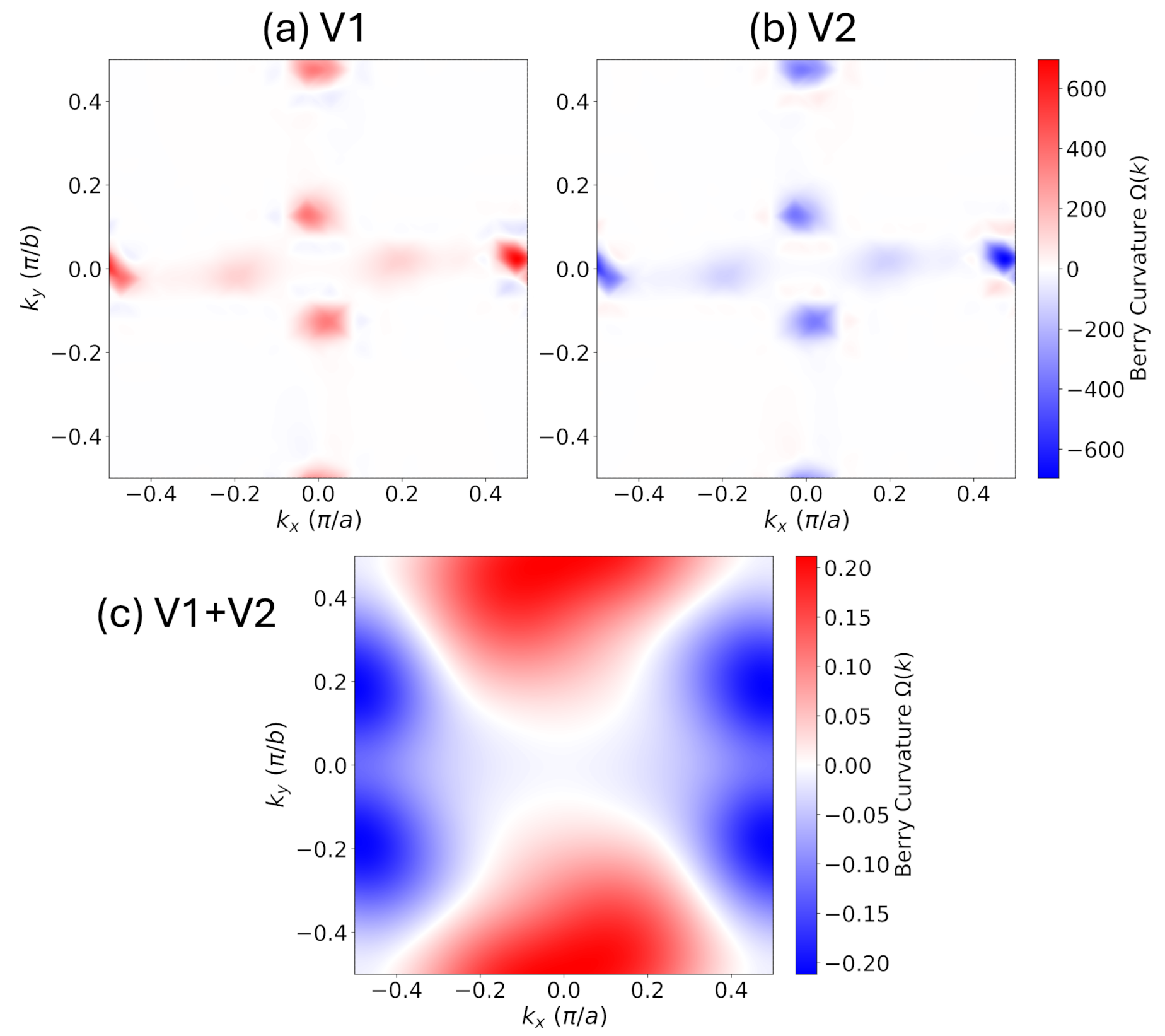} 
\caption{%
Momentum-space distribution of the Berry curvature $\Omega(\mathbf{k})$ for the top two valence bands (V1 and V2) of monolayer AgF$_2$, corresponding to the spin-orbit coupled band structures shown in Fig.~\ref{fig:bands-soc}(b-d).
\textbf{(a)}~Berry curvature of the V1 band, which yields a Chern number of $C^E = +3$.
\textbf{(b)}~Berry curvature of the V2 band, with an opposite Chern number of $C^E = -3$.
\textbf{(c)}~Total Berry curvature, $\Omega_{\mathrm{V1}} + \Omega_{\mathrm{V2}}$, exhibiting strong local hotspots but a vanishing net Chern number.}
\label{fig:berry-ele}
\end{figure*}

\subsection{Multipolar Analysis}

To quantify the momentum-space symmetry of the spin splitting, we perform a multipolar expansion~\cite{leon2025strain,fukaya2025superconducting}, which distinguishes between odd-parity ($p$-wave) and even-parity ($d$-wave) character. The detailed methodology of this expansion is provided in the Supplemental Material. In the absence of spin-orbit coupling, AgF$_2$ behaves as a pure $d$-wave altermagnet, with an even-parity response constrained by symmetry.

The inclusion of SOC introduces relativistic interactions that activate odd-parity ($p$-wave) components, transforming the system into a hybrid $d/p$-wave state. Quantitatively, the overall $d$-wave sector remains slightly dominant ($f_d \approx 0.53$ vs. $f_p \approx 0.47$), but the admixture is strongly anisotropic: the $y$ ($f_p \approx 0.57$) and $z$ ($f_p \approx 0.60$) channels acquire a pronounced $p$-wave character, while the $x$ channel remains predominantly $d$-wave ($f_d \approx 0.65$).

This relativistic $d/p$ admixture underpins the dual character of the tilted polar phase: a robust $d$-wave background that preserves the compensated collinear order, and a subdominant $p$-wave fingerprint along $M'$--$\Gamma$--$M$ that drives nonreciprocal effects, spin-galvanic couplings~\cite{hellenes2023p, song2025electrical}, and DMI-like interactions shaping both electronic and magnonic dispersions. In contrast to canonical $p$-wave magnets~\cite{hellenes2023p, fukaya2025superconducting}, this hybridization arises without non-collinear textures, establishing relativistic altermagnetism as the organizing principle of AgF$_2$’s unique symmetry fingerprints.


\subsection{AM+SOC: Topological potential}
To investigate the topological character of the altermagnetic phase, we compute the Berry curvature $\Omega_n(\mathbf{k})$ for the two highest valence bands (V1 and V2), as shown in Fig.~\ref{fig:berry-ele}. Integration over the Brillouin zone yields the corresponding Chern number for each band.
Our analysis reveals a remarkable topological potential in the altermagnetic phase. Upon inclusion of spin-orbit coupling (SOC), the top valence band, V1 (higher band at $M'$), acquires a large Chern number of $C^E_1 = +3$, driven by pronounced hotspots of positive Berry curvature. In direct contrast, the adjacent band V2 carries an exactly opposite charge of $C^E_2 = -3$, with its curvature distribution mirroring that of V1. As a result, the total Chern number of the occupied valence manifold is $C^E_{\text{tot}} = C^E_1 + C^E_2 = 0$.

To overcome this perfect cancellation of topological charges, we suggest three pathways to engineer a net topological response. The goal is to apply a perturbation that affects V1 and V2 differently, thus breaking their compensation. For example, a uniaxial strain or a perpendicular electric field could break the symmetries that protect the V1-V2 degeneracy, opening a selective energy gap and isolating the contribution of a single band\cite{gonzalez2025altermagnetism,huang2024strain,you2021electric,chen2024strain,bao2025isolated}. An alternative approach involves using a patterned substrate, heterostructures, or site-selective doping to create a staggered potential, which would shift the energy of one band relative to the other\cite{hadad2016self,lian2025tunable,li2023quantum}. Furthermore, designing a heterostructure that hybridizes the AgF$_2$ monolayer with another material could selectively modify one of the valence bands, leaving the topological character of the other intact. Implementing any of these perturbations could transform the system into an intrinsic Chern insulator, characterized by a quantized anomalous Hall conductance $\sigma_{xy} = (e^2/h) C^E_{\text{tot}}$ with $|C^E_{\text{tot}}| = 3$. This tunable topology highlights the potential of altermagnetic materials as switchable platforms for topological transport phenomena.

\begin{figure*}[!]
\centering
\includegraphics[clip,width=0.7\textwidth]{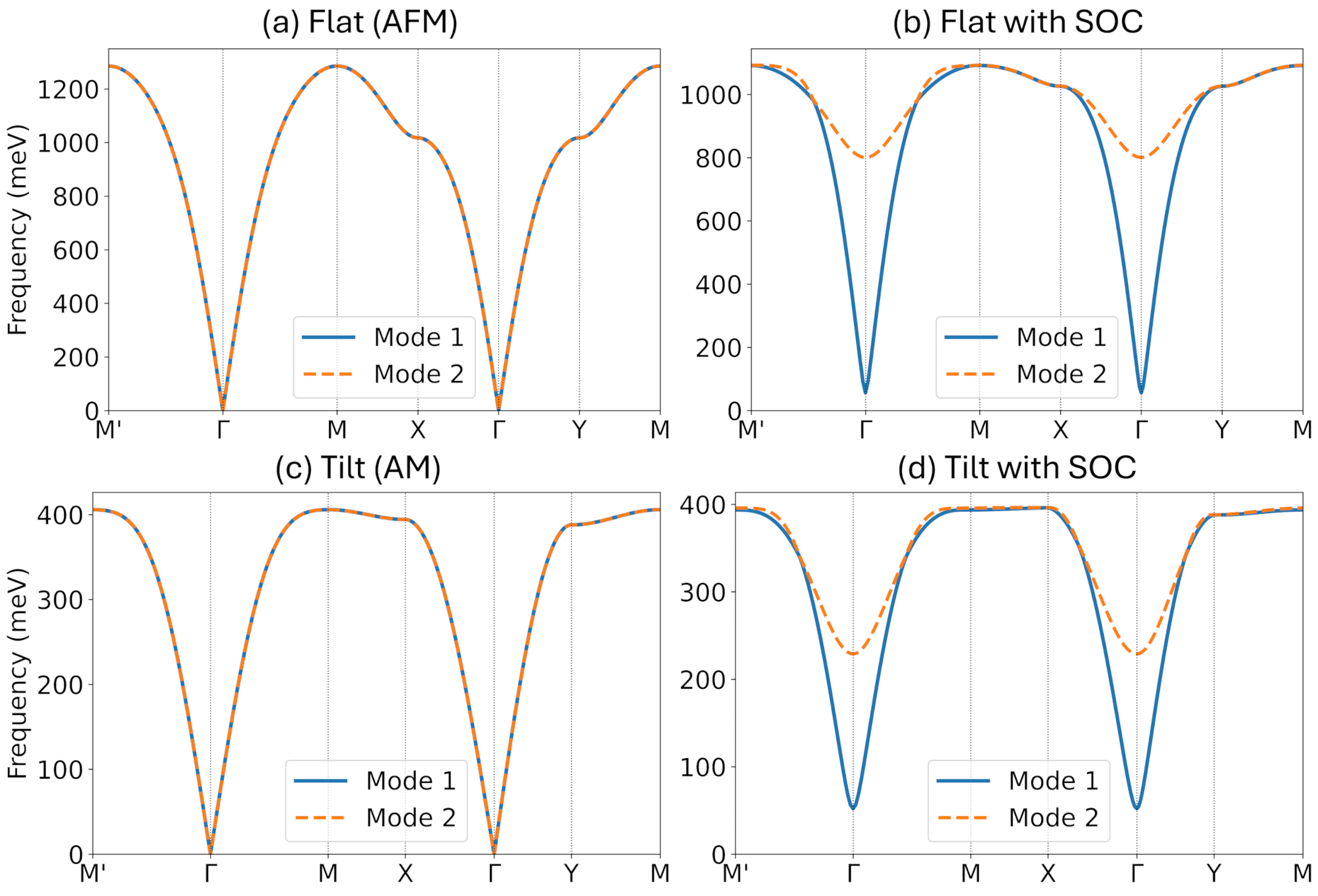}
\caption{%
Magnon spectra of monolayer AgF$_2$ in the flat antiferromagnetic phase [top row, panels~\textbf{(a)} and~\textbf{(b)}] and the tilted altermagnetic  phase [bottom row, panels~\textbf{(c)} and~\textbf{(d)}]. 
\textbf{(a, c)}~Spectra computed without spin-orbit coupling; 
\textbf{(b, d)}~corresponding spectra with SOC included.}
\label{fig:magnon_bands}
\end{figure*}

While such high Chern numbers have been reported in ferromagnetic 2D systems such as CoBr$_2$ \cite{chen2017intrinsic}, MoF$_3$, and members of the WX$_3$ family \cite{bao2025isolated}, their realization in antiferromagnetic systems typically requires non-collinear spin textures or additional symmetry-breaking mechanisms \cite{lian2020flat}. In contrast, AgF$_2$ achieves this topological regime through the interplay of spin-orbit coupling and its intrinsic altermagnetic symmetry. Although SOC is essential to activate the Berry curvature, the symmetry of the non-relativistic altermagnetic phase defines the structure of the bands and enables their topological character upon turning on relativistic effects.
These findings position AgF$_2$ as a promising candidate for realizing topological quantum phases in magnetically compensated, spin-orbit-active materials.

\section{Topological Magnons Driven by Altermagnetic Symmetry}

In the insulating regime of monolayer AgF$_2$, charge excitations are suppressed by a large electronic band gap of approximately 1.8~eV. As a result, the low-energy physics is governed by collective spin dynamics, i.e., magnons, rather than electronic quasiparticles. Having established that the structural distortion to the polar $P2_1$ space group breaks global inversion symmetry in the altermagnetic ground state, we now investigate how this reduced symmetry imprints itself on the magnon spectrum. The absence of inversion symmetry enables momentum-dependent spin splitting in the electronic bands, raising the question of whether a similar mechanism leads to a topological magnon phase.

To address this, we construct an effective spin Hamiltonian with isotropic Heisenberg exchange, Dzyaloshinskii–Moriya interaction, and symmetric anisotropic exchange parameters extracted from first-principles DFT+U calculations~\cite{ozaki2004numerical,ozaki2005efficient,he2021tb2j}. Within linear spin-wave theory\cite{MAGNOPY,ivanov2021fast}, we compute the magnon modes and analyze their topological character. The mangnon band topology is quantified by the integer Chern number ($C^M$), obtained from the Berry curvature of the Bogoliubov-de Gennes eigenvectors~\cite{shindou2013topological} using a gauge-invariant numerical scheme~\cite{fukui2005chern}. The detailed formalism is provided in the Supplemental Material.

A nonzero Chern number signals a topological phase, which manifests as a finite transverse thermal Hall conductivity ($\kappa_{xy}$)~\cite{Katsura2010,onose2010observation}. In this regime, the Berry curvature acts as an effective magnetic field in momentum space, deflecting magnon flow under a longitudinal temperature gradient, producing a transverse thermal Hall current, $\kappa_{xy}$, and, since magnons carry spin angular momentum, an accompanying spin current, phenomena closely related to the spin Seebeck and spin Nernst effects~\cite{uchida2008observation}. Our calculations of $\kappa_{xy}$ and the associated Chern numbers thus predict measurable transverse heat and spin transport in finite samples, establishing transverse thermal transport as a direct probe of altermagnet-induced magnon topology in AgF$_2$~\cite{nagaosa2013topological,uchida2014longitudinal}.


\subsection{Altermagnons on AgF$_2$}
We now apply the methodology established in the preceding sections to the specific case of monolayer AgF$_2$. Our analysis demonstrates how the intrinsic altermagnetic symmetry of this material gives rise to a topological magnon phase. A direct comparison between the high-symmetry, flat antiferromagnetic phase and the distorted, tilted altermagnetic state reveals the critical role of symmetry in enabling this phenomenon.

\begin{table}[h!]
\centering
\caption{Dominant exchange parameters for the flat, centrosymmetric AFM phase, computed without and with spin-orbit coupling (SOC). All values are in meV. Note that DMI is symmetry-forbidden without SOC and remains negligible when SOC is included.}
\label{tab:afm_params}
\begin{tabular}{l| c | c c}
\hline
\hline
Bond & $J_{\mathrm{iso}}$ & $J_{\mathrm{iso}}$ & $|\mathbf{D}|$ \\

 & (no SOC) &\multicolumn{2}{c}{(with SOC)} \\
\hline
$J_1$ & -17.44 & -16.99 & $2 \times 10^{-4}$ \\
$J_2$ & +6.52  & +3.16  & 0 \\
$J_3$ & -0.18  & -0.18  & 0 \\
\hline
\hline
\end{tabular}
\end{table}

\subsubsection{The Topologically Trivial Antiferromagnetic Phase}
We analyze the flat, centrosymmetric AFM phase ($P4/mmm$ space group) as the high-symmetry reference for our study. In the absence of spin-orbit coupling, the extracted spin Hamiltonian is purely isotropic and thus exhibits full SU(2) spin-rotation invariance. The spontaneous breaking of this continuous symmetry by the Néel order generates gapless magnon modes at the $\Gamma$ point, consistent with Goldstone’s theorem. Moreover, the $P4/mmm$ symmetry enforces degeneracy of the two magnon modes across the Brillouin zone, as shown in Fig.~\ref{fig:magnon_bands}(a).

Including SOC does not alter this picture significantly, since the high crystal symmetry largely suppresses relativistic effects. As reported in Table~\ref{tab:afm_params}, the DMI interaction, a key ingredient for magnon topology, remains negligible ($|D|/|J_1| \sim 10^{-4}$). SOC does break the SU(2) spin-rotation invariance, lifting the Goldstone protection and opening a small anisotropy gap at $\Gamma$ [Fig.~\ref{fig:magnon_bands}(b)]. Nevertheless, the Berry curvature remains identically zero throughout the Brillouin zone, confirming that the magnon bands in this phase are topologically trivial. 

Beyond these essential features, the AFM magnon spectrum displays a remarkably large bandwidth. The dominant nearest-neighbor coupling, $J_1 \approx -17\,$meV, produces a total dispersion exceeding $1.2\,$eV [Fig.~\ref{fig:magnon_bands}(a)], unusually large for a 2D antiferromagnet. In the long-wavelength limit near $\Gamma$, the acoustic branch follows the linear dispersion $\hbar\omega(\mathbf{k}) \approx v_s |\mathbf{k}|$, where the high spin-wave velocity $v_s$ reflects the rigidity of the spin medium. The $P4/mmm$ symmetry also forbids anisotropic interactions, thereby enforcing exact degeneracy of the two branches across the Brillouin zone. While the Mermin–Wagner theorem would preclude long-range order in a strictly isotropic 2D system, the small anisotropy gap induced by SOC [Fig.~\ref{fig:magnon_bands}(b)] is sufficient to stabilize the AFM order. Altogether, these results establish the flat AFM phase as an ideal non-topological baseline for assessing how tilt distortions and SOC drive AgF$_2$ into an altermagnetic topological magnon phase.


\begin{figure}[!]
\centering
\includegraphics[width=0.5\textwidth,trim=0pt 0pt 25pt 0pt,clip]{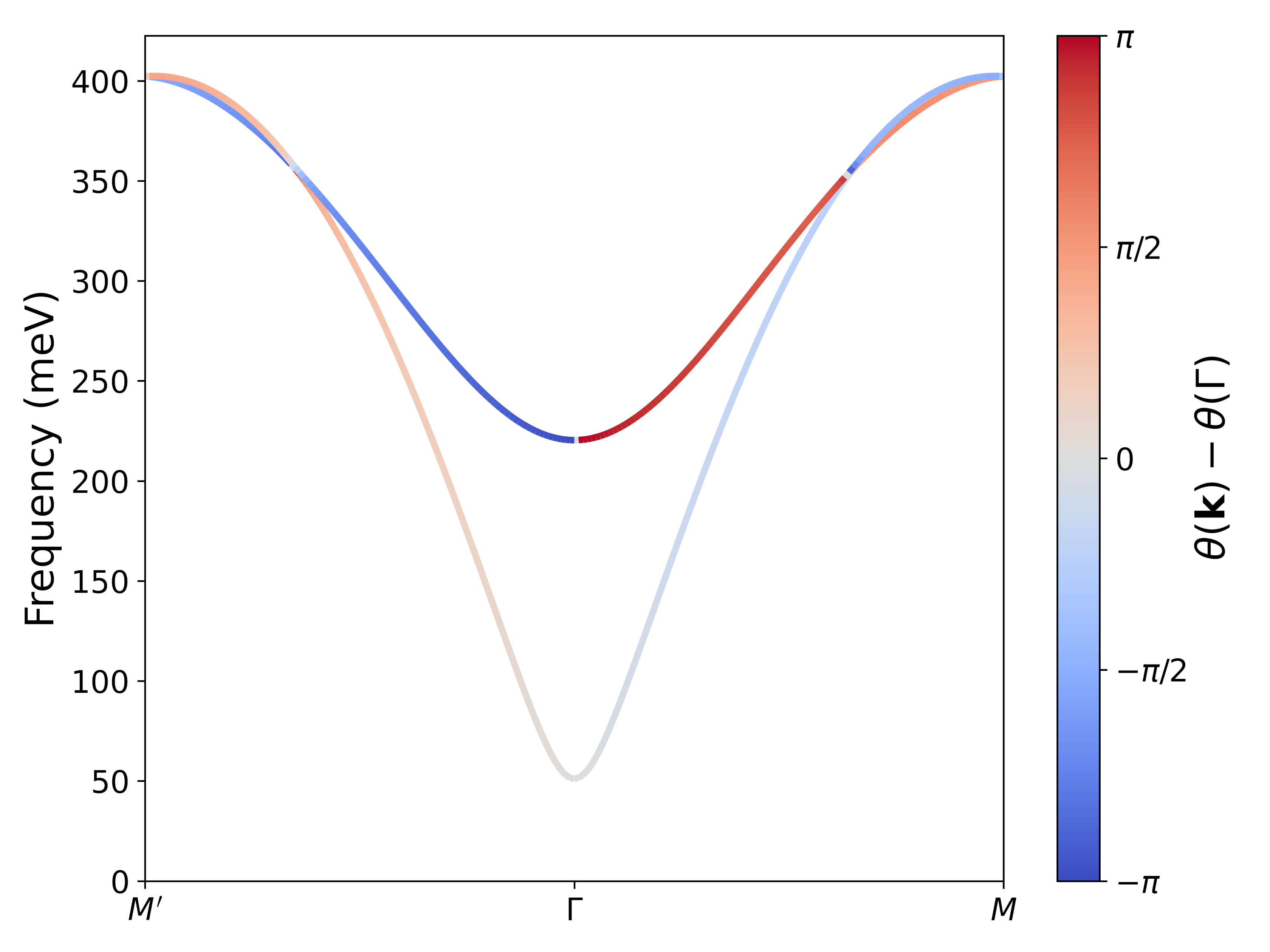}
\caption{
Chirality of the magnon modes in the altermagnetic (AM) phase of monolayer AgF$_2$, computed along the high-symmetry path $\mathrm{M}'$-$\Gamma$-$\mathrm{M}$.
The color scale represents the relative phase angle of magnon precession,  $\varphi_s(\mathbf{k}) - \varphi_s(\mathbf{\Gamma})$, which quantifies the momentum-dependent chirality of the two non-degenerate modes. }
\label{fig:magnon_quiral}
\end{figure}

\begin{figure*}[!]
\centering
\includegraphics[clip,width=0.8\textwidth]{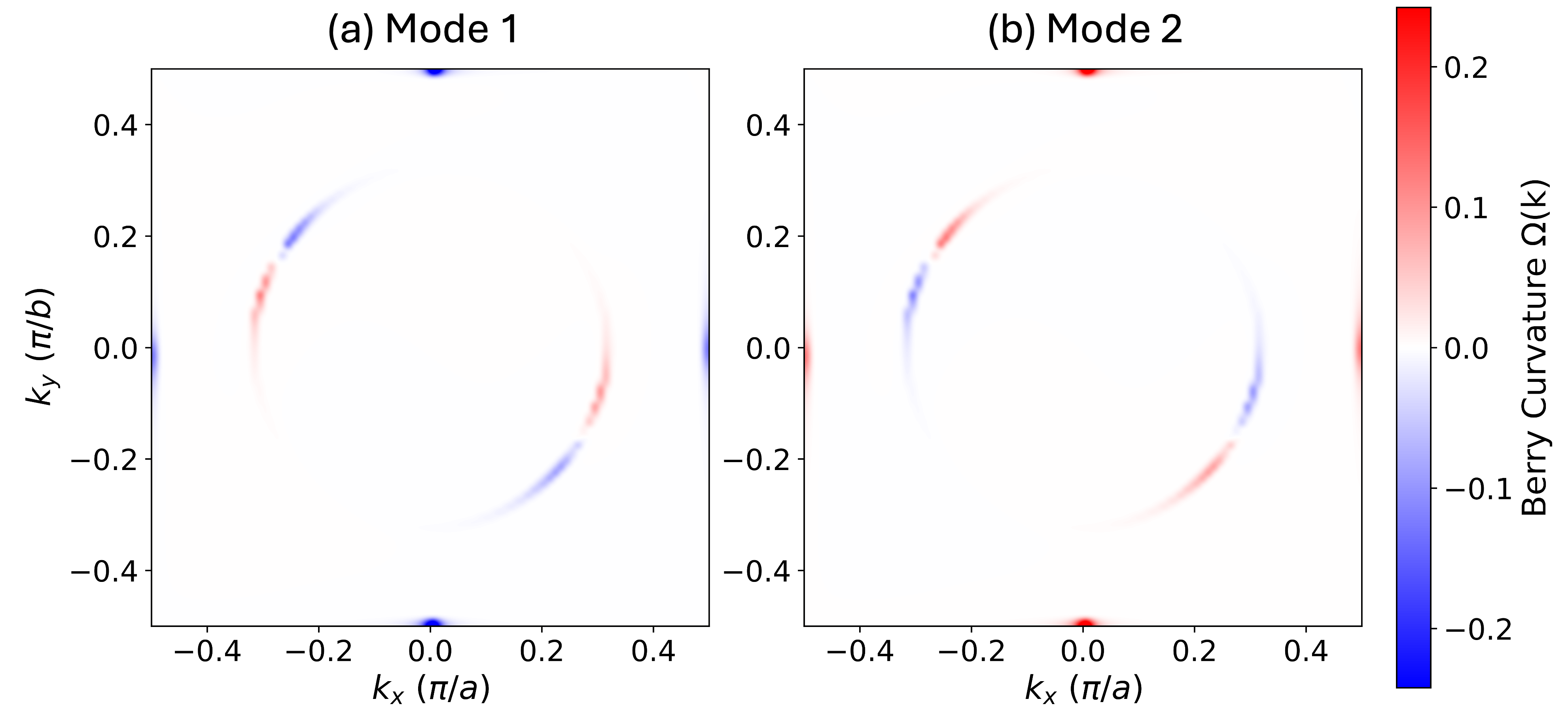}
\caption{%
Momentum-space distribution of the Berry curvature $\Omega(\mathbf{k})$ for the two lowest-energy magnon modes in the altermagnetic (AM) phase of monolayer AgF$_2$ with spin-orbit coupling (SOC).
\textbf{(a)}~Berry curvature of Mode 1, which yields a Chern number of $C^M = +1$.
\textbf{(b)}~Berry curvature of Mode 2, with an opposite topological charge and a Chern number of $C^M = -1$.}
\label{fig:magnon_berry}
\end{figure*}

\subsubsection{Symmetry-Driven Topology in the Altermagnetic Phase}

Unlike the trivial AFM phase, the tilted altermagnetic ground state (polar $P2_1$ space group) provides the symmetry breaking required for non-trivial magnon topology. The structural distortion removes the inversion center and reshapes the magnetic exchange network by weakening $J_1$, reversing the sign of $J_2$, and collapsing the magnon bandwidth by more than 60\%. As detailed in Table~\ref{tab:am_params_extended}, all DMI terms vanish in the absence of SOC, and the spectrum exhibits gapless Goldstone modes at $\Gamma$, confirming that the distortion alone cannot induce a topological phase. Nevertheless, it generates strongly momentum-dependent spin textures and creates ideal conditions for relativistic effects. The resulting magnon modes acquire opposite chirality, directly reflecting the altermagnetic symmetry and establishing the key prerequisite for topology, as confirmed by their spin precession~\cite{hoyer2025altermagnetic}.

\begin{table}[h!]
\centering
\caption{Dominant exchange parameters ($>0.1$\,meV) for the tilted altermagnetic  phase, without and with spin-orbit coupling. The total magnitudes of the DMI interaction ($|\mathbf{D}|$) and the symmetric anisotropic exchange ($|J_{\mathrm{ani}}|$, Frobenius norm) are shown. All values are in meV.}
\label{tab:am_params_extended}
\begin{tabular}{c | c | c c c}
\hline
\hline
Bond & $J_{\mathrm{iso}}$ & $J_{\mathrm{iso}}$ & $|\mathbf{D}|$ & $|J_{\mathrm{ani}}|$ \\
& (no SOC) & \multicolumn{3}{c}{(with SOC)}  \\
\hline
$J_1$ & -9.27 & -10.66 & 1.55 & 0.14 \\
$J_2$ & -9.28 & -10.69 & 1.99 & 0.31 \\
$J_3$ & -9.28 & -10.69 & 1.99 & 0.31 \\
$J_4$ & -9.57 & -10.61 & 1.55 & 0.14 \\
\hline
$J_5$ & +1.05 & +1.08  & $\sim0$ & 1.68 \\
$J_6$ & +0.88 & +0.89  & $\sim0$ & 1.74 \\
\hline
$J_7$ & +0.28 & +0.32  & $\sim0$ & 0.53 \\
$J_8$ & -0.10 & -0.10  & 0.03 & $\sim0$ \\
\hline
\hline
\end{tabular}
\end{table}

When SOC enters, anisotropic interactions dominated by a strong Dzyaloshinskii-Moriya interaction (DMI) give the magnon bands a topological character. Calculations of the Berry curvature reveal sharp enhancements near the $X$ and $Y$ points, where inter-band gaps narrow and spin textures wind rapidly. Integration over the Brillouin zone yields quantized Chern numbers, $C^M_1=+1$ and $C^M_2=-1$, for the two lowest bands. The finite Berry curvature produces a transverse thermal Hall conductivity ($\kappa_{xy}$), a hallmark of topological magnon bands~\cite{Katsura2010}. The low symmetry of the $P2_1$ phase permits this response, and the calculated magnitude matches values reported in pyrochlore and kagome magnets~\cite{onose2010observation}, suggesting that experiments should access it.

The thermal Hall signal [Fig. \ref{fig:magnon_cond}] should appear in the low-to-intermediate temperature range below the estimated Néel temperature, $T_N \sim 100$–$150$ K, where the altermagnetic order and topological bands remain stable. At higher temperatures, magnon–magnon and magnon–phonon scattering will likely reduce the response~\cite{hoyer2025altermagnetic}, but the low-temperature regime firmly establishes the tilted phase of AgF$_2$ as an altermagnetic topological magnon insulator. Unlike higher-symmetry altermagnets where thermal Hall effects are forbidden, or hematite where topology relies on long-range exchange~\cite{hoyer2025altermagnetic}, AgF$_2$ derives its topological character from a strong nearest-neighbor DMI enabled directly by a polar structural distortion. This mechanism opens a distinct route to robust topological phenomena in collinear, compensated magnets.



\section{Final Remarks}

A central finding of this work is the discovery of a dual topological phase in monolayer AgF$_2$, where non-trivial character emerges in both its electronic and magnonic excitations, driven by the same underlying physical mechanism. This dual nature stems from a symmetry-lowering structural distortion to a polar, altermagnetic ground state. By breaking inversion symmetry, this distortion provides the essential platform for relativistic spin-orbit coupling to become active, which is otherwise suppressed in the high-symmetry, trivial antiferromagnetic phase.
This symmetry-driven mechanism has profound and parallel consequences. For the electronic structure, it enables a non-relativistic altermagnetic spin splitting and, once SOC is included, gives rise to a remarkable topological state where the two highest valence bands acquire large and opposite integer Chern numbers of $C^E=\pm3$. For the magnons, the same mechanism activates a strong nearest-neighbor DMI interaction. This, in turn, opens a topological gap in the magnon spectrum, creating two chiral bands with robust, quantized Chern numbers of $C^M=\pm1$.
The non-trivial bulk topology of these magnonic bands has direct, measurable consequences. It produces a finite transverse thermal Hall conductivity, $\kappa_{xy}$, providing a macroscopic fingerprint of the non-zero Berry curvature. Furthermore, the bulk-boundary correspondence dictates that these Chern numbers imply the existence of unidirectional, counterpropagating, dissipationless magnon edge modes~\cite{hatsugai1993chern}. While our linear spin-wave theory faithfully describes the low-temperature regime, we anticipate that at higher temperatures ($T \gtrsim T_N/5$), scattering from magnon-magnon and magnon-phonon interactions will cause spectral broadening and potentially affect the topological signatures
~\cite{Mook2021}.
Finally, the coexistence of topological magnon bands ($C^M=\pm1$) with electronic bands carrying higher Chern number ($C^E=\pm3$) identifies monolayer AgF$_2$ as a promising platform to investigate the interplay between topological magnons and electrons. Resonant inelastic x-ray scattering, sensitive to both charge and spin excitations, could directly reveal signatures of interconversion and hybrid modes in this system~\cite{Li2023_RIXS}.

Equally important is the broader conceptual message. The relativistic altermagnetic state of AgF$_2$ naturally hosts a hybrid parity structure: a dominant even-parity, $d$-wave-like spin splitting complemented by a subdominant, SOC-enabled odd-parity $p$-wave component. This $d/p$ admixture is a unique fingerprint of relativistic altermagnetism, providing the microscopic pathway to the observed dual electronic and magnonic topology. At the same time, the stabilization of chiral magnon bands with quantized Chern numbers constitutes a direct realization of altermagnonics, the topological spin dynamics of relativistic altermagnets. Taken together, our findings show that altermagnets are not only fertile ground for electronic topology but also natural cradles of bosonic topology. This establishes a new arena where fermionic and bosonic quasiparticles are intrinsically entangled, highlighting relativistic altermagnetism as a unifying framework for multifunctional quantum matter and pointing toward future device concepts based on lattice, spin, and symmetry engineering.

\section*{Acknowledgments}
The authors acknowledge the financial support provided by FONDECYT Regular Grants 1250364 and 1250803, as well as the Basal Program for Centers of Excellence through Grant CIA250002 (CEDENNA). Powered@NLHPC: This research was partially supported by the supercomputing infrastructure of the NLHPC (CCSS210001).

\section*{Competing Interests}
The Authors declare no Competing Financial or Non-Financial Interests.

\section*{{Credit} authorship contribution statement}
J.W.G. and A.L. carried out the electronic analysis; J.W.G., R.A.G. and N.V.S. performed the magnonic study; all authors contributed to the design, discussion of results, and writing of the manuscript.

\section*{Data Availability}
The data that support the findings of this study are available 
from the corresponding author, upon reasonable request.

\bibliography{suya}

\newpage
\pagebreak
\newpage
\onecolumngrid
\begin{center}

\Large{Supplementary Information}

\Large{Dual Topology as a Fingerprint of Relativistic Altermagnetism in AgF$_2$ Monolayer}

\end{center}

\setcounter{figure}{0} 
\setcounter{section}{0} 
\setcounter{equation}{0}
\setcounter{table}{0}
\setcounter{page}{1}
\renewcommand{\thepage}{S\arabic{page}} 
\renewcommand{\thesection}{S\Roman{section}}   
\renewcommand{\thetable}{S\arabic{table}}  
\renewcommand{\thefigure}{S\arabic{figure}} 
\renewcommand{\theequation}{S\arabic{equation}} 

\section{Symmetry Analysis of the Distorted Phase}

To confirm the loss of inversion symmetry in the distorted altermagnetic phase, we performed a symmetry analysis using \textsc{FINDSYM}~\cite{stokes2005findsym,FINDSYM} with strict tolerances ($10^{-4}$ \AA{} for the lattice and $10^{-3}$ \AA{} for atomic positions). The resulting space group is $P2_1$ (No.~4), a non-centrosymmetric polar group that allows spin-orbit interactions. This structure emerges from a symmetry-lowering distortion of the high-symmetry tetragonal phase $P4/mmm$ (No.~123), in which cooperative octahedral tilting and buckling break both inversion and fourfold rotational symmetry. 

Interestingly, the centrosymmetric phase $P2_1/c$ (No.~14) is not a subgroup of $P4/mmm$, but it is a maximal supergroup of $P2_1$, related through an index-2 subgroup relation. This suggests that a $P2_1/c$-like polymorph, characterized by octahedral tilting without buckling, could represent a nearby metastable structure. While it is not crystallographically connected to the tetragonal parent, it may still belong to the broader symmetry-lowering landscape. Thus, the observed $P2_1$ ground state could arise either directly from $P4/mmm$ or through an intermediate step involving a $P2_1/c$-like configuration, which provides a useful reference for describing the underlying structural order parameter.

In the distorted $P2_1$ phase, the original inversion center is replaced by a two-fold screw axis. This $2_1$ operation combines a 180$^\circ$ rotation around the crystallographic $b$-axis with a translation of half a lattice vector along the same axis. It maps one Ag sublattice onto the other, so the two spin-opposed sublattices are no longer related by inversion but by a nonsymmorphic screw symmetry. This fulfills the defining symmetry condition for altermagnetism~\cite{Smejkal2022_Review} and, crucially, enables spin-orbit coupling to generate Dzyaloshinskii-Moriya interactions. These relativistic terms are strictly forbidden in the centrosymmetric $P4/mmm$ phase but are essential for stabilizing the topological magnon bands observed in the altermagnetic state.

\section{Spin-orbit coupling effects in the altermagnetic phase}

\subsection{Symmetry analysis}

To clarify the role of relativistic spin–orbit coupling (SOC) in the distorted altermagnetic phase of AgF$_2$, we analyze the magnetic space group symmetry in detail. The tilted $P2_1$ structure lowers the symmetry relative to the tetragonal $P4/mmm$ parent, removing inversion and fourfold rotational symmetry. The relevant magnetic space group is $P2_1'$ (No.~4.9), which contains the anti-unitary operation $(C_{2b}|\tfrac{1}{2}\mathbf{b})\mathcal{T}$: a two-fold screw rotation combined with time reversal. In momentum space, this operation maps $\mathbf{k} \to -C_{2b}\mathbf{k}$ and $\mathbf{S}\to -\mathbf{S}$, thereby enforcing degeneracies only along specific high-symmetry lines such as $\Gamma$--$Y$ ($k_x=0$). Away from these lines (e.g. $M'$--$\Gamma$--$M$), no such constraint holds, and exchange-driven, even-parity ($d$-wave–like) altermagnetic splittings are symmetry-allowed~\cite{Smejkal2022_Review}.

Moreover, SOC lifts residual degeneracies at zone-boundary points. Most notably, gaps of up to 20~meV open at $M$ and $M'$ [Fig.~\ref{fig:bands-flat-tilt}(c)]. Using the \textsc{MAXMAGN} tool~\cite{MAXMAGN} of the Bilbao Crystallographic Server~\cite{aroyo2006bilbao,bilbao}, we find that the little group at $M=(\tfrac{1}{2},\tfrac{1}{2},0)$ is the type-I magnetic space group $P_s1$ (No.~1.3), which lacks anti-unitary operations. In particular, the symmetry $(C_{2b}|\tfrac{1}{2}\mathbf{b})\mathcal{T}$ does not leave $M$ invariant, since $C_{2b}\mathbf{k}\not\equiv\mathbf{k}$ modulo a reciprocal lattice vector. As a result, no symmetry protects degeneracy at $M$ and $M'$, and SOC opens sizable gaps at these points.

\subsection{Magnetic anisotropy and weak ferromagnetism}

SOC also determines the magnetic ground state. Our calculations yield a magnetic anisotropy energy (MAE) of $\sim$0.2~meV per formula unit, favoring the $xy$ plane over the $z$ axis. With spins aligned along $y$, the low-symmetry crystal field activates Dzyaloshinskii–Moriya interactions (DMI) and anisotropic exchanges, producing a small canting of the antiparallel spins. This results in a net weak ferromagnetic moment of $\sim$0.1~$\mu_B$, primarily along $x$. Such canting is symmetry-allowed in the polar $P2_1$ phase, and vanishes in the centrosymmetric flat polymorph $P4/mmm$, where global inversion strictly forbids DMI~\cite{moriya1960anisotropic}. Our calculations confirm vanishing net moments ($<10^{-4}\,\mu_B$) in the flat structure.

The emergence of weak ferromagnetism has several implications: (i) it couples the AM order to external magnetic fields, (ii) DMI generates nonreciprocal magnon dispersions $\omega(\mathbf{k}) \neq \omega(-\mathbf{k})$, relevant for magnonic circuits~\cite{jungwirth2025altermagnetism}, and (iii) the net moment provides an experimental probe through spin-torque or magneto-optical techniques.

\subsection{Spin texture and parity mixing}

Finally, we analyze the spin polarization of the top valence bands (V1, V2) in Fig.~\ref{fig:bands-soc}. In the non-relativistic limit [Fig.~\ref{fig:bands-soc}(a)], the splitting is purely $d$-wave–like and collinear. Inclusion of SOC produces a non-collinear spin texture, with $\langle S_y \rangle$ dominant but finite $\langle S_x \rangle$ and $\langle S_z \rangle$ components that vary across the Brillouin zone. Along $\Gamma$--$Y$, the anti-unitary screw symmetry enforces degeneracy, but along generic paths like $M'$--$\Gamma$--$M$, no such constraint exists, and SOC reveals richer anisotropies.

Beyond opening band gaps, SOC introduces an odd-parity component to the spin splitting. This $p$-wave–like channel coexists with the dominant $d$-wave response, generating a hybrid altermagnetic state with expanded symmetry-breaking phenomenology~\cite{fukaya2025superconducting,wang2025two,leon2025strain}. Consequences include anisotropic transport responses, directional dichroism, spin-galvanic effects, and enhanced tunability of the relativistic altermagnetic phase.

\section{Multipolar analysis: Technical details}

\subsection{Formal definitions}

To resolve the symmetry character of the spin-resolved splitting $\Delta E^\beta(\mathbf{k})$ ($\beta\in\{x,y,z\}$), we perform a multipolar decomposition in terms of Brillouin-zone (BZ) moments of the in-plane momentum $\mathbf{k}_\parallel=(k_x,k_y)$~\cite{leon2025strain,fukaya2025superconducting}.  
The first two multipoles are defined as
\begin{align}
P_{\alpha}{}^{\beta} &= \Big\langle k_\alpha\,\Delta E^\beta(\mathbf{k}) \Big\rangle, \\
Q_{\alpha\gamma}{}^{\beta} &= \Big\langle k_\alpha k_\gamma\,\Delta E^\beta(\mathbf{k}) \Big\rangle,
\end{align}
with $\alpha,\gamma\!\in\!\{x,y\}$.  
Odd dependence in $\mathbf{k}$ corresponds to $p$-wave character, while even dependence corresponds to $d$-wave character.  
The BZ average is defined as
\begin{equation}
\langle f\rangle \equiv \Big(\sum_{\mathbf{k}} w_{\mathbf{k}}\Big)^{-1} \sum_{\mathbf{k}} w_{\mathbf{k}}\, f(\mathbf{k}),
\end{equation}
where $w_{\mathbf{k}}$ are the $k$-point integration weights from the self-consistent calculation.

For each spin channel $\beta$, we define $L^1$ norms of the dipolar and quadrupolar tensors:
\begin{equation}
M_p^\beta=\sum_\alpha \big|P_\alpha^{\ \beta}\big|,\qquad
M_d^\beta=\sum_{\alpha,\gamma}\big|Q_{\alpha\gamma}^{\ \ \beta}\big|.
\end{equation}
From these we obtain the fractional contributions,
\begin{equation}
f_p^\beta=\frac{M_p^\beta}{M_p^\beta+M_d^\beta},\qquad
f_d^\beta=\frac{M_d^\beta}{M_p^\beta+M_d^\beta}.
\end{equation}
Overall fractions are reported as $f_p=\sum_\beta f_p^\beta$ and $f_d=1-f_p$.

\subsection{Collinear case}

In the collinear limit, the spin quantization axis is global and the splitting is extracted from the up- and down-spin bands of the occupied set $\mathcal{O}$:
\begin{equation}
\Delta E^z(\mathbf{k}) = \frac{1}{|\mathcal{O}|} \sum_{b\in\mathcal{O}} \!\Big[E_{b,\uparrow}(\mathbf{k}) - E_{b,\downarrow}(\mathbf{k})\Big].
\end{equation}
This construction captures the even-in-$\mathbf{k}$ splitting enforced by symmetry, which corresponds to a purely $d$-wave altermagnetic character.

\subsection{Inclusion of spin-orbit coupling}

When SOC is included, Bloch states become spinors with band-resolved spin expectation $\mathbf{S}_n(\mathbf{k})$. To obtain a symmetry-faithful splitting we construct a spin-contrast projection vector from selected band pairs $(i,j)$:
\begin{equation}
\Delta E_{ij}^\beta(\mathbf{k}) = \Big[E_i(\mathbf{k})-E_j(\mathbf{k})\Big]\, u_{ij}^\beta(\mathbf{k}),
\end{equation}
\begin{equation}
\mathbf{u}_{ij}(\mathbf{k}) \equiv \frac{\mathbf{S}_i(\mathbf{k})-\mathbf{S}_j(\mathbf{k})}
{\big\|\mathbf{S}_i(\mathbf{k})-\mathbf{S}_j(\mathbf{k})\big\|+\epsilon},
\end{equation}
where $\epsilon>0$ prevents singularities near avoided crossings.  
The effective splitting is then the arithmetic mean over the selected pairs,
\begin{equation}
\Delta E^\beta(\mathbf{k}) = \frac{1}{|\mathcal{O}|}\sum_{(i,j)\in\mathcal{P}} \Delta E_{ij}^\beta(\mathbf{k}),
\end{equation}
with $\mathcal{O}$ denoting the chosen set of near-Fermi bands.


In the non-relativistic case, the tilted polar phase of AgF$_2$ exhibits a robust $d$-wave altermagnetic response, dominated by the quadrupolar tensor $Q^z$.  
This reflects its even-parity, exchange-driven splitting.  
The inclusion of SOC activates additional odd-parity ($p$-wave) components, producing a  hybrid $d/p$ state.  
Quantitatively, we find $f_d \approx 0.53$ and $f_p \approx 0.47$, indicating nearly balanced contributions.  
The admixture is anisotropic, where the $x$ channel remains dominantly $d$-wave ($f_d \approx 0.65$),   while the $y$ ($f_p \approx 0.57$) and $z$ ($f_p \approx 0.60$) channels are strongly $p$-wave.  
The dominant $d$-wave sector maintains compensated collinear order, while the emergent $p$-wave component governs relativistic responses such as nonreciprocal transport, spin-galvanic effects~\cite{hellenes2023p, song2025electrical}, and linear DMI-like terms in both electronic and magnonic dispersions.  
The coexistence of these channels establishes AgF$_2$ as a relativistic altermagnet with dual $d/p$ fingerprints, in contrast to canonical $p$-wave magnets that require non-collinear spin textures to preserve time-reversal symmetry~\cite{hellenes2023p,fukaya2025superconducting}.  
This relativistic admixture provides the microscopic pathway to intertwined electronic and magnonic topology, and is consistent with recent proposals of $p$-wave channels mediating superconductivity~\cite{sukhachov2025coexistence} and their electrical control~\cite{song2025electrical}.

\section{Formalism for Topological Magnon Calculations}

\subsection{Spin Hamiltonian and Linear spin-wave theory }
We model the low-energy spin excitations in AgF$_2$ using a classical spin Hamiltonian, with parameters extracted from first-principles DFT+$U$ calculations performed with the \texttt{OpenMX} package~\cite{ozaki2004numerical,ozaki2005efficient}. The \texttt{TB2J} package~\cite{he2021tb2j} is used to obtain the isotropic Heisenberg couplings $J^{\rm iso}_{ij}$, the Dzyaloshinskii–Moriya interaction (DMI) vectors $\mathbf{D}_{ij}$, and the symmetric anisotropic exchange tensors $\mathbf J_{ij}^{\rm ani}$ for the Hamiltonian:
\begin{equation}
 \mathcal{H} = \sum_{i<j}J^{\rm iso}_{ij}\,\mathbf S_i\cdot\mathbf S_j
+ \sum_{i<j}\mathbf D_{ij}\cdot(\mathbf S_i\times\mathbf S_j)
+ \sum_{i<j}\mathbf S_i\cdot\mathbf J_{ij}^{\rm ani}\cdot\mathbf S_j.
\end{equation}
Linear spin-wave theory  is then applied using the \texttt{Magnopy} package\cite{MAGNOPY,ivanov2021fast}, which constructs and diagonalizes the bosonic Bogoliubov–de Gennes (BdG) Hamiltonian, $\mathcal{H}_{\text{BdG}}(\mathbf{k})$, yielding the magnon dispersion $\omega_n(\mathbf{k})$ and eigenvectors. To characterize the modes, we compute the spin expectation value $\langle \mathbf{S}_n(\mathbf{k}) \rangle$ for each band $n$ using the particle-particle ($U(\mathbf{k})$) and particle-hole ($V(\mathbf{k})$) components of the Bogoliubov transformation matrix:
\begin{align}
\langle S^x_n(\mathbf{k})\rangle &= \Re\sum_{i=1}^{M}U_{i n}^*(\mathbf{k})\,V_{i n}(\mathbf{k}), \\
\langle S^y_n(\mathbf{k})\rangle &= \Im\sum_{i=1}^{M}U_{i n}^*(\mathbf{k})\,V_{i n}(\mathbf{k}), \\
\langle S^z_n(\mathbf{k})\rangle &= \sum_{i=1}^{M}\Bigl(\lvert U_{i n}(\mathbf{k})\rvert^2 - \lvert V_{i n}(\mathbf{k})\rvert^2\Bigr).
\end{align}

\subsection{Topological Characterization}
We assess the magnon band topology by computing the Chern number of each band, $C_n$, defined as the integral of the Berry curvature $\Omega_n(\mathbf k)$ over the Brillouin Zone (BZ):
\begin{equation}
C_n = \frac{1}{2\pi}\!\int_{\text{BZ}}\!\Omega_n(\mathbf k)\,d^2k.
\end{equation}
The Berry curvature is derived from the BdG eigenvectors $\ket{\psi_n(\mathbf k)}$ with the bosonic paraunitary normalization $\langle\psi_n|\tau_z|\psi_n\rangle=1$~\cite{shindou2013topological}. Numerically, we implement the gauge-invariant Fukui scheme~\cite{fukui2005chern} on a discrete $\mathbf{k}$-mesh. The Berry flux $F_{12}(\mathbf{k})$ for each plaquette is calculated as:
\begin{equation}
\begin{split}
F_{12}(\mathbf{k}) = \arg \Biggl[ & \langle \psi_n(\mathbf{k}) | \psi_n(\mathbf{k}+\hat{k}_x) \rangle \langle \psi_n(\mathbf{k}+\hat{k}_x) | \psi_n(\mathbf{k}+\hat{k}_x+\hat{k}_y) \rangle \\
& \langle \psi_n(\mathbf{k}+\hat{k}_x+\hat{k}_y) | \psi_n(\mathbf{k}+\hat{k}_y) \rangle \langle \psi_n(\mathbf{k}+\hat{k}_y) | \psi_n(\mathbf{k}) \rangle \Biggr].
\end{split}
\end{equation} 
The total Chern number is the sum over all plaquettes:
\begin{equation}
C^M = \frac{1}{2\pi}\sum_{\mathbf k}F_{12}(\mathbf k).
\end{equation} 

\subsection{Thermal Hall Conductivity}
The transverse thermal Hall conductivity, $\kappa_{xy}$, is calculated to quantify the topological response~\cite{Katsura2010,onose2010observation}, using the formula:
\begin{equation}
\kappa_{xy}(T)
= -\frac{k_B^2T}{\hbar V}\sum_{n, \mathbf{k}}
c_2\bigl[\rho(\omega_n(\mathbf{k}))\bigr]\,\Omega_n(\mathbf{k}),
\end{equation} 
where $\rho$ is the Bose–Einstein distribution and $c_2[x] = (1+x)\ln(1+x) - x\ln(x)$ is the standard entropic weight function~\cite{Katsura2010}.  

Together, this workflow provides the formalism used to obtain the magnon spectra, Berry curvature, Chern numbers, and thermal Hall response reported in the main text.

\begin{figure}[!]
\centering
\includegraphics[clip,width=0.5\textwidth]{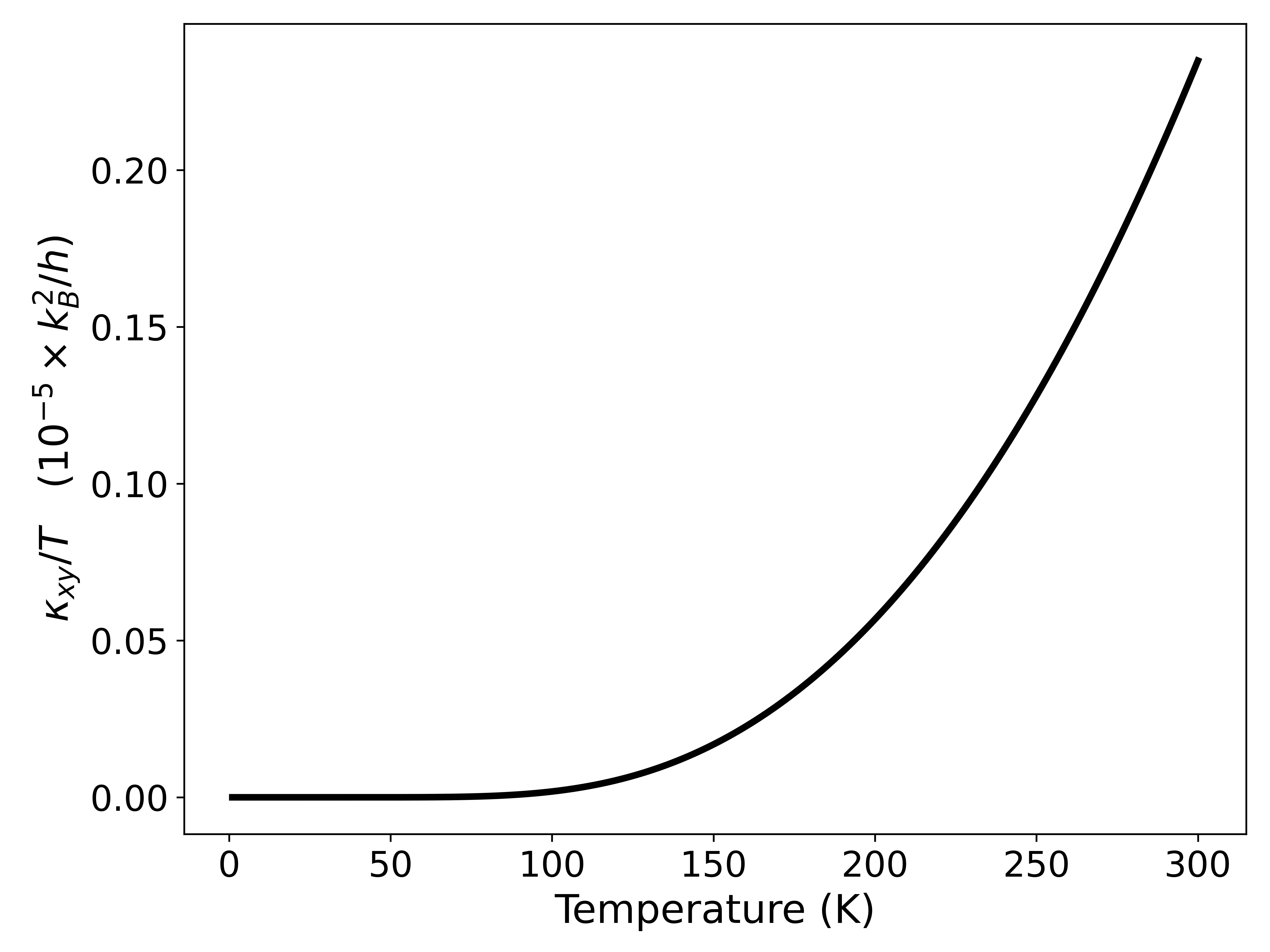}
\caption{%
Temperature dependence of the magnon thermal Hall conductivity, $\kappa_{xy}/T$, in the altermagnetic (AM) phase of monolayer AgF$_2$.
This quantity characterizes the transverse heat current carried by magnons and reflects the topological nature of the magnon bands.
The conductivity vanishes at low temperatures and increases with thermal population of the magnon modes.}
\label{fig:magnon_cond}
\end{figure}

\section{Momentum-space spin texture}
\begin{figure*}[!]
\centering
\includegraphics[clip,width=0.8\textwidth]{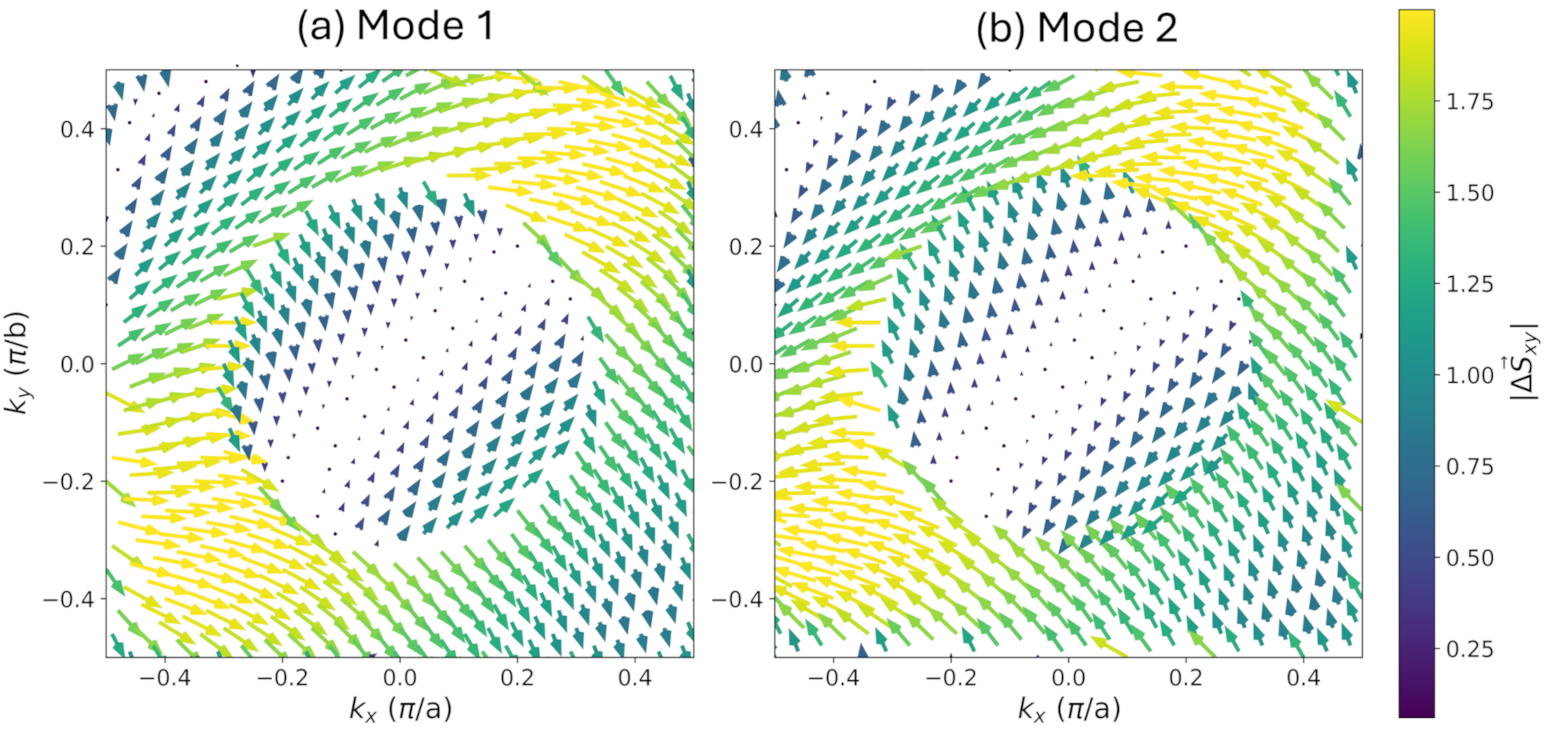}
\caption{ Momentum-space spin texture, $\langle \vec{S}_n(\mathbf{k}) \rangle$, for the two lowest-energy magnon modes, Mode 1 \textbf{(a)} and Mode 2 \textbf{(b)}, in the altermagnetic (AM) phase of monolayer AgF$_2$ with spin-orbit coupling (SOC).
The arrows represent the direction of the in-plane spin polarization, $\vec{S}_{xy} = (S_x, S_y)$, while the color scale indicates its magnitude.
The distinct vortex-like winding patterns reveal the opposite chirality of the two modes, arising from their nontrivial topological character and associated Chern numbers.}
\label{fig:magnon_texture}
\end{figure*}

The microscopic origin of the nontrivial topology is most clearly revealed by the momentum-space spin textures, $\langle\vec{S}_n(\mathbf{k})\rangle$, shown in Fig.~\ref{fig:magnon_texture}. These vector fields depict the in-plane spin components, $\mathbf{S}_{xy} = (S_x, S_y)$, and display distinct vortex-like winding patterns centered at the $\Gamma$ point. As expected, the two magnon modes exhibit opposite chiralities: Mode~1 (panel a) winds counterclockwise, while Mode~2 (panel b) winds clockwise.
This winding directly encodes the topological character of the bands. The local spin texture governs the Berry curvature via the expression
\begin{equation}
\Omega_n(\mathbf{k}) \propto \hat{S}_n \cdot \left( \partial_{k_x} \hat{S}_n \times \partial_{k_y} \hat{S}_n \right),
\end{equation}
where $\hat{S}_n = \langle \vec{S}_n \rangle / |\langle \vec{S}_n \rangle|$ is the normalized spin vector. Regions of rapid momentum-space spin rotation yield large contributions to $\Omega_n(\mathbf{k})$. The winding can also be quantified via the so-called skyrmion density in momentum space,
\begin{equation}
\rho_{\mathrm{sk},n}(\mathbf{k}) = \frac{1}{4\pi} \hat{S}_n(\mathbf{k}) \cdot \left( \partial_{k_x} \hat{S}_n(\mathbf{k}) \times \partial_{k_y} \hat{S}_n(\mathbf{k}) \right),
\end{equation}
whose Brillouin zone integral yields the same Chern number derived from the Berry curvature. For the textures shown, the counterclockwise vortex of Mode~1 corresponds to $C^M_1 = +1$, while the clockwise vortex of Mode~2 yields $C^M_2 = -1$.
Thus, Fig.~\ref{fig:magnon_texture} provides a direct and intuitive visualization of the topological invariant: the opposite chiralities of the two magnon modes are not merely geometric features, they are the momentum-space signature of their quantized topological character.

\end{document}